\documentclass[sn-mathphys]{sn-jnl}

\jyear{2022}

\newcommand{\epem}     {\ensuremath{\mathrm{e^+e^-}}}

\newcommand{\znull}    {\ensuremath{\mathrm{Z^0}}}
\newcommand{\mz}     {\ensuremath{m_{\znull}}}
\newcommand{\mh}     {\ensuremath{m_H}}
\newcommand{\mb}     {\ensuremath{m_{\mathrm{b}}}}
\newcommand{\mbos}   {\ensuremath{m_{\mathrm{b}}^{\mathrm{OS}}}}
\newcommand{\asmz}   {\ensuremath{\as(\mz)}}
\newcommand{\mbmz}   {\ensuremath{\mb(\mz)}}
\newcommand{\mbmh}   {\ensuremath{\mb(\mh)}}
\newcommand{\mbmb}   {\ensuremath{\mb(\mb)}}
\newcommand{\msbar}  {\ensuremath{\overline{\mathrm{MS}}}}
\newcommand{\bbbar}  {\ensuremath{\mathrm{b\overline{b}}}}
\newcommand{\ffbar}  {\ensuremath{\mathrm{f\overline{f}}}}
\newcommand{\qqbar}  {\ensuremath{\mathrm{q\overline{q}}}}
\newcommand{\rb}     {\ensuremath{\mathrm{R_{0,b}}}}
\newcommand{\as}     {\ensuremath{\alpha_{\mathrm{S}}}}
\newcommand{\roots}  {\ensuremath{\sqrt{s}}}

\begin{document}

\title[\mbmz\ revisited with Zedometry]{\mbmz\ revisited with Zedometry}

\author[1]{\fnm{S.}\sur{Kluth}}
\email{skluth@mpp.mpg.de}

\affil[1]{
  \orgname{Max-Planck-Institut f\"ur Physik},
  \orgaddress{
    \street{F\"ohringer Ring 6},
    \postcode{80805}, \city{Munich}
    \country{Germany}
  }
}

\abstract{Precision measurements of \znull\ boson properties could
  enable a determination of the mass of the b quark at the scale of
  the \znull\ boson mass \mbmz. The dependence of Standard Model
  predictions on the b quark mass using the program Gfitter is
  studied.  The precision of the currently available measurements by
  the LEP experiments and SLD, together with measurements from the LHC
  experiments of the mass of the top quark and the Higgs boson, is not
  sufficient for a relevant determination. The predicted precision of
  \znull\ boson resonance measurements at future \epem\ colliders will
  allow a competitive direct determination of \mbmz.}


\maketitle

\section{Introduction}

The theory of strong interactions, Quantum Chromo Dynamics
(QCD)~\cite{fritzsch73,gross73a,gross73b,politzer73} is a part of the
Standard Model (SM) of particle physics via QCD corrections to all
electroweak interactions~\cite{glashow61,weinberg67,salam79} involving
quarks. QCD has as the only free parameters the strong coupling
constant, usually given as its value at the \znull\ mass scale \asmz,
and the masses of the six quarks of the SM. The values of the strong
coupling as well as of the quark masses depend on and decrease with
the energy scale of the interaction, which is known as asymptotic
freedom.

The mass of the b quark in the \msbar\ renormalisation
scheme~\cite{msbar} was determined from analysis of B-hadron mass
spectra and thus at energy scales corresponding to the B-hadron masses
with the current world average
$\mbmb=(4.183\pm0.004)$~GeV~\cite{pdg20}.  Measurements of the mass of
the b quark at the energy scale of the \znull\ boson mass \mbmz\ were
performed by the LEP experiments using jet production with events
tagged by B-hadron decays and next-to-leading order QCD
calculations. A review~\cite{kluth06} quotes $\mbmz=(2.90\pm0.31)$~GeV
where the error is dominated by experimental and hadronisation
systematic uncertainties. These results for the b quark mass in the
\msbar\ scheme are different by more than four standard deviations and
thus already provide strong evidence for the presence of a running b
quark mass. This finding was recently reproduced and
improved~\cite{aparisi21}.

The new analysis~\cite{aparisi21} studied the dependence of the
branching ratio of Higgs boson decays to a \bbbar\ pair
$\Gamma(H\rightarrow\bbbar)$ normalised to the branching ratio for
Higgs boson decays to a pair of \znull\ bosons.  The branching ratio
$\Gamma(H\rightarrow\bbbar) \sim \mbmh^2$ and therefore a measurement
is highly sensitive to \mbmh. The analysis obtains
$\mbmh=(2.60^{+0.36}_{-0.31})$~GeV from combined recent ATLAS and CMS
measurements of
$\Gamma(H\rightarrow\bbbar)/\Gamma(H\rightarrow\znull\znull)$, which
has the same precision as the combined LEP measurements of \mbmz.

We propose to study the dependence of \znull\ boson properties
connected with b quarks on the value of the b quark mass \mbmz\ used
in a SM prediction. The observables are the partial width
$\Gamma(Z\rightarrow\bbbar)$, the branching ratio
$BR(Z\rightarrow\bbbar)=\Gamma(Z\rightarrow\bbbar)/\Gamma_{Z,tot}$,
where $\Gamma_{Z,tot}$ is the total width of the \znull\ boson, and
$\rb=\Gamma(Z\rightarrow\bbbar)/\Gamma(Z\rightarrow hadrons)$.  A
comparison of these predictions with precision measurements of these
quantities can be used to extract \mbmz.

A determination of \mbmh\ from $\Gamma(H\rightarrow\bbbar)$ assumes
the Yukawa coupling of the Higgs boson to b quarks to have the
expected SM value~\cite{bi20,aparisi21}. A complementary determination
of the b quark mass at high scales, i.e.\ \mbmz, but without direct
dependence on the b quark Yukawa coupling, will thus help to resolve
possible ambiguities.

We will briefly review the dependence of SM predictions for
\znull\ boson properties on \mbmz, then explain the programs used to
obtain predictions, compare the predictions to measurements with
uncertainties as valid now and with uncertainties expected for the
proposed future facility FCC-ee~\cite{fcc18}.

\section{SM predictions as a function of $m_b$}

The SM prediction for $\Gamma(Z\rightarrow\ffbar)$, where \ffbar\ is a
pair of SM fermions, can be written as
\begin{equation}
  \Gamma(Z\rightarrow\ffbar) = \frac{G_f\mz^3}{24\sqrt{2}\pi} N_{C,f}
  (v_f^2R_f^V+a_f^2R_f^A)  
\end{equation}
where $G_F$ is the Fermi constant, $N_{C,f}$ is the number of colours,
and $v_f$ and $a_f$ are the vector and axial-vector couplings of the
\znull\ to the fermions. The radiator functions $R_f^V$ and $R_f^A$
contain in the case of a \qqbar\ pair the QCD corrections, see
e.g.~\cite{chetyrkin93}.  The leading QCD corrections to the radiator
functions for $\znull\rightarrow\bbbar$ decays\footnote{The fermion
  index $f$ is dropped now for clarity.} are
\begin{equation}
  R^{V/A}(s) = 1 + R_1^{V/A}\frac{\as}{\pi} + \left(\frac{\mb(s)}{Q}\right)^2
  \left( R_0^{V/A(m)} + R_1^{V/A(m)}\frac{\as}{\pi} \right) + \ldots
\end{equation}
where $Q=\roots$. The mass corrections up to $\as^3(\mb/Q)^4$ and
$\as^3(\mb/Q)^2$ are known~\cite{chetyrkin00}.

The observables depend on the b quark mass $\sim-\mb^2$ from the
leading $R_0^{A(m)}$ mass correction term~\cite{chetyrkin93} and thus
one expects that predictions for the three observables will decrease
quadratically with increasing \mbmz.  In comparison to
$\Gamma(H\rightarrow\bbbar)$ the sensitivity of
$\Gamma(Z\rightarrow\bbbar)$ to $\mb(s)^2$ is suppressed by about a
factor $s=\mz^2$.

Predictions of the SM for \znull\ boson properties are available in
the programs zfitter~\cite{zfitter,zfitter2} and
Gfitter~\cite{gfitter08}.  The last available version 6.42 of the
zfitter program dates from 2006 and does not contain updates to SM
predictions which appeared since this date. The Gfitter program was
used for the last update of the electroweak precision
fit in 2018~\cite{gfitter18}. We use Gfitter version
2.2~\footnote{http://project-gfitter.web.cern.ch/project-gfitter/Software/Gfitter2\_2.tar.bz2}
and for comparison and cross checks zfitter version
6.42~\footnote{https://elsevier.digitalcommonsdata.com/895c9aea-260c-4c90-91ee-261e32e04f19}.

\subsection{Gfitter predictions}
\label{sec_gfitter}

The Gfitter program uses \mbmb\ in the \msbar\ scheme as input with a
default value of $\mbmb=4.2$~GeV. The Gfitter parameter \verb+mb_MSb+
is defined in the Gfitter datacard with a scan range from 1.0 to
6.0~GeV.  We set in section \verb+GEWFlags+ of the Gfitter datacard
the flag \verb+FullTwoLoop = "F"+, which turns this option off,
because the full two-loop prediction in Gfitter is a parametrisation
without variation of \mbmb.  Using the Gfitter action
\verb+FctOfFreePara =+ \verb+"T:mb_MSb:Nbins=50"+ the parameter
\verb+mb_MSb+ is varied over its range with 50 points and the
predictions for the active theory parameters are calculated. As active
theory parameters we define \verb+GEW::R0b+, \verb+GEW::GammaZhad+ and
\verb+GEW::GammaZtot+, where we implemented the class \verb+GammaZhad+
in the \verb+GEW+ package of Gfitter to provide the prediction for
$\Gamma(\znull\rightarrow hadrons)$. Using these three predicted
parameters we can derive the observables $\Gamma(Z\rightarrow\bbbar)$
and $BR(Z\rightarrow\bbbar)$. The scan of \verb+mb_MSb+ is repeated
for three values of \asmz\ given by $\asmz=0.1179\pm0.0010$.  For the
evaluation of theory uncertainties the N3LO terms $C_{04}\as^4$ and
$I_4\as^4$ in the radiator functions are multiplied simultaneously by
factors of zero or two with the Gfitter parameters
\verb+DeltaAlphasTheoC05_Scale+ and
\verb+DeltaAlphasTheoCMt4_Scale+~\footnote{In order to avoid double
  counting we disabled separate variation of the results for \rb,
  $\Gamma_{Z,tot}$ and $\Gamma_{Z,had}$ with the same parameters in
  Gfitter.} and the scans of \verb+mb_MSb+ are repeated.

\subsection{zfitter predictions}

The zfitter program implements heavy quark masses in the on-shell (OS)
scheme using a value of $\mbos=4.7$~GeV. The on-shell mass is
converted in zfitter to the running mass in the \msbar\ scheme and
then evolved to the \znull\ mass scale. The zfitter program calculates
and prints the predictions of \znull\ properties such as
$\Gamma(\znull\rightarrow\bbbar)$ via the routine \verb+ZVWEAK+. The
value of the input OS b quark mass $\mbos$ is varied over a range
from 1.0 to 7.0~GeV in steps of 0.2~GeV. The scan of $\mbos$ is
repeated for three values of \asmz\ given by $\asmz=0.1179\pm0.0010$.

\subsection{Mass scheme conversion and mass evolution}

In Gfitter the \msbar\ mass with a nominal value of $\mbmb=4.2$~GeV is
used as input. We use 4-loop evolution of \msbar\ quark masses to
present Gfitter results at the \znull\ mass scale. For all
computations the same value of $\asmz=0.1179\pm0.0010$ as above is
applied.  The calculations available in
CRunDec~\cite{rundec00,crundec12,crundec17} are used for all mass
transformations~\footnote{https://github.com/DavidMStraub/rundec-python}. For
estimating systematic uncertainties we change the value of
\asmz\ inside its errors, or change the number of loops, i.e.\ the
perturbative accuracy, from the standard value of four to three.

In zfitter the OS mass \mbos\ is used as an input parameter.  In order
to present results at the \znull\ mass scale we convert from the OS to
the \msbar\ mass definition at the scale of \mbos\ using the 4-loop
calculations implemented in CRunDec. The resulting $\mb(\mbos)$ value
is evolved at 4-loop accuracy to \mbmz.

\section{Results}

The LEP and SLD results for the three observables are
$\Gamma(\znull\rightarrow\bbbar) = 377.3 \pm 1.2$~MeV,
$BR(Z\rightarrow\bbbar) = 15.121 \pm 0.048$~\% and $\rb = 0.21628 \pm
0.00066$~\cite{zedometry05}. The values and errors shown are from the
determinations assuming lepton universality. The relative total
uncertainties are 0.3\% for all three observables.

The predictions by Gfitter are presented below. All plots on
figure~\ref{fig_gfitter} show as solid black lines the Gfitter
predictions for the observables as indicated as a function of
\mbmz. The expected decrease of the observables $\sim(\mbmz)^2$ is
clearly visible.

\begin{figure}[htb!]
  \begin{center}
    \includegraphics[width=\textwidth]{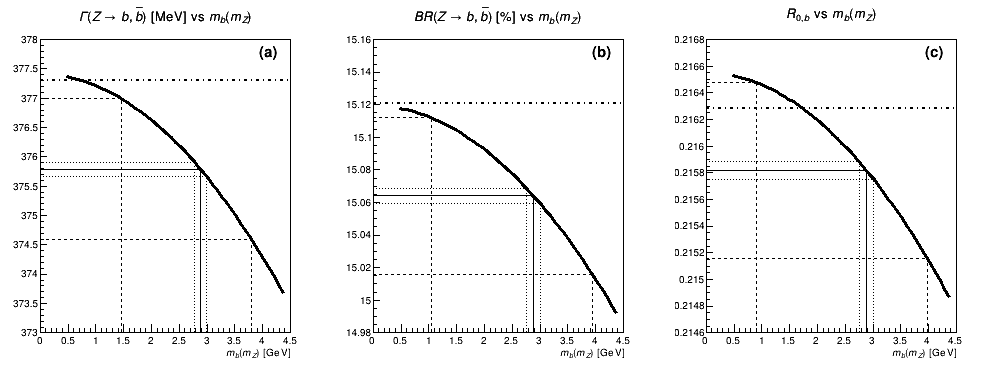}
    \caption[bla]{Figures (a), (b) and (c) show Gfitter predictions
      for the three observables as a function of \mbmz. The horizontal
      dash-dotted lines indicate the measurements from LEP and SLD.
      The horizontal and vertical solid and dashed lines display the
      correspondence between a hypothetical determination based on the
      Gfitter default value $\mbmz=2.88$~GeV ($\mbmb=4.2$~GeV), and
      results for \mbmz. The dotted lines show the same correspondence
      for experimental uncertainties reduced by a factor 1/10 (see
      text for details).}
    \label{fig_gfitter}
  \end{center}
\end{figure}

On all plots of figure~\ref{fig_gfitter} the horizontal dash-dotted
lines indicate the LEP and SLD results~\cite{zedometry05}.  On
figures~\ref{fig_gfitter} (a), (b) and (c) the horizontal and vertical
solid lines indicate the value of the observable leading to a result
$\mbmz=2.88$~GeV ($\mbmb=4.2$~GeV), i.e.\ the nominal Gfitter
value. The horizontal dashed lines indicate the actual uncertainties
from the LEP and SLD results applied to the nominal value calculated
by Gfitter and the vertical dashed lines thus indicate the
corresponding uncertainty region for \mbmz.  The uncertainties of a
hypothetical determination of $\mbmz=2.88$~GeV have values between
approximately one and two GeV.

The actual LEP and SLD results with their uncertainties cannot be used
to derive a determination of \mbmz, since already their one s.d.\ upper
uncertainty intervals cover regions not accessible by the predictions.
The incompatibility is less pronounced for \rb, where possible biases
to the hadronic and \bbbar\ widths of the \znull\ boson can cancel
more effectively in the ratio.

Comparing the three observables we find that the uncertainties of the
hypothetical determinations of \mbmz\ increase from
$\Gamma(\znull\rightarrow\bbbar)$ to \rb. This can be explained by the
fact that $BR(Z\rightarrow\bbbar)$ and \rb\ are normalised by the
total or hadronic width of the \znull, which also depend on
\mb\ leading to a reduction of the experimental sensitivity.

The horizontal dotted lines in figures~\ref{fig_gfitter} (a), (b) and
(c) present the same hypothetical determination as above with total
uncertainties taken as 1/10th of the actual uncertainties. The studies
in~\cite{fcc18} for possible precision measurements of \znull\ boson
properties at FCC-ee show that such a precision could be reached. The
resulting uncertainties for a possible determination of \mbmz\ are
shown by the dotted vertical lines. We conclude that with measurements
of the observables with such small uncertainties precise
determinations of \mbmz\ can be obtained.

With a numerical evaluation of the hypothetical determination of
\mbmz\ we study theoretical uncertainties. The value of
$\asmz=0.1179\pm 0.0010$ is varied within its uncertainties for the
Gfitter predictions. The results are shown in
table~\ref{tab_gfitter}. The theory uncertainty (see
section~\ref{sec_gfitter}) is presented in the last column labeled ``N3LO
unc.''.

\begin{table}[htb!]
  \begin{center}
    \begin{tabular}{lclll}
      Observable & expected value & \mbmz & \mbmz & \mbmz \\ & &
      exp. unc. & \asmz\ unc. & N3LO unc. \\ \hline
      $\Gamma(\znull\rightarrow\bbbar)$ [MeV] & $375.8 \pm 1.2$ &
      $_{-1.41}^{+0.92}$ & $\pm 0.09$ & $\pm 0.01$ \\ & $375.79 \pm
      0.12$ & $^{+0.10}_{-0.11}$ & $\pm 0.09$ & $\pm0.01$ \\ \hline
      $BR(Z\rightarrow\bbbar)$ [\%] & $15.064 \pm 0.048$ &
      $_{-1.84}^{+1.06}$ & $\pm 0.03$ & $\pm0.02$ \\ & $15.0641 \pm
      0.0048$ & $_{-0.13}^{+0.12}$ & $\pm 0.03$ & $\pm0.02$ \\ \hline
      \rb & $0.21582 \pm 0.00066$ & $_{-1.98}^{+1.09}$ & $\pm 0.01$ &
      $\pm0.04$ \\ & $0.215818 \pm 0.000066$ & $\pm 0.13$ & $\pm 0.01$
      & $\pm0.04$ \\
    \end{tabular}
    \caption{Results for \mbmz\ uncertainties [GeV] with Gfitter for
      the three observables with either the actual uncertainties from
      LEP and SLD or uncertainties reduced by a factor 1/10.
      $\mbmz=2.88$~GeV is used in all cases.}
    \label{tab_gfitter}
  \end{center}
\end{table}

We find that an assumed reduction of the experimental uncertainties
from LEP and SLD by a factor 1/10 leads to a corresponding reduction
of the experimental uncertainty of \mbmz\ by about the same
factor. Furthermore, as observed above, the experimental uncertainties
increase slightly from $\Gamma(\znull\rightarrow\bbbar)$ to \rb\ also
with the reduced measurement errors. However, the uncertainty of the
predictions due to variation of $\asmz=0.1179\pm0.0010$ is smaller for
$BR(Z\rightarrow\bbbar)$ w.r.t.\ $\Gamma(\znull\rightarrow\bbbar)$ and
is negligible for the determination of \mbmz\ derived from a
measurement of \rb. This effect was already discussed for
\rb\ in~\cite{chetyrkin96}.  The relative total uncertainty on
\mbmz\ with reduced uncertainties is about 5\%.

The N3LO theory uncertainties are smaller than the experimental
uncertainties with the expected measurements at a future FCC-ee. The
N3LO theory uncertainties are smaller than the \asmz\ uncertainty for
$\Gamma(\znull\rightarrow\bbbar)$, they are about the same size for
$BR(Z\rightarrow\bbbar)$, and for \rb\ they are larger than the
\asmz\ uncertainty.

The b quark mass is set in Gfitter as \mbmb\ which is a technical
limitation in the analysis, since this value must be evolved to \mbmz,
even though the SM prediction could be used directly with \mbmz\ as a
free parameter. As a test of the additional uncertainty introduced by
the evolution from \mbmb\ to \mbmz\ the value of
$\asmz=0.1179\pm0.0010$ is changed within its errors and the
perturbative order of the calculation is changed from 4-loop to 3-loop
precision. The corresponding uncertainties for \mbmz\ are
$_{-0.04}^{+0.03}$ (\asmz) and $\pm0.10$ (pert. order). Since these
uncertainties would not appear if \mbmz\ could be varied directly we
do not show them in table~\ref{tab_gfitter}.

It was checked that the results for the uncertainties are consistent
with the results from zfitter.

\section{Conclusions}

We have studied the possibility to determine \mbmz\ from precision
measurements of \znull\ boson properties from LEP and SLD or the
proposed future \epem\ facility FCC-ee. The theory predictions for the
three observables $\Gamma(\znull\rightarrow\bbbar)$,
$BR(Z\rightarrow\bbbar)$, and \rb\ are obtained from the program
Gfitter as functions of the b quark mass \mbmz, evolved from the
Gfitter input parameter \mbmb.  A comparison of the predictions with
actual measurements from LEP and SLD does not provide a meaningful
extraction of \mbmz.  Using expected central values with current
uncertainties corresponding to the input b quark mass of the
predictions we find that with the data from LEP and SLD the
uncertainties for \mbmz\ from the analysis of \znull\ boson properties
would be between one and two GeV. These uncertainties are much larger
than other existing determinations of \mbmz\ from jet production in
b-tagged events at LEP or \mbmh\ from Higgs boson decays to b quark
pairs.

Using the expected uncertainties for measurements of \znull\ boson
properties at the future \epem\ collider FCC-ee, which are taken to be
smaller by a factor 1/10 w.r.t.\ the current uncertainties, a
determination of \mbmz\ with a relative error of 5\% is possible.  The
total uncertainty would be dominated by the experimental errors in all
cases.  Depending on the observable either the uncertainty from the
current world average value of \asmz\ or from the N3LO theory error
would be the second largest uncertainty.

At future \epem\ colliders determinations of \mbmh\ with O(10)~MeV
precision could be obtained~\cite{aparisi21}. The uncertainty for a
determination of \mbmz\ from jet production in b-tagged hadronic final
states in \epem\ annihilation at the proposed ILC based on a large
sample of \znull\ boson decays was estimated as
$\Delta\mbmz=0.12$~GeV~\cite{ildphyspub2021001}. This uncertainty is
dominated by theory uncertainties. Our proposed determination of
\mbmz\ is thus expected to have a compareable uncertainty with
complementary sources of uncertainties.

A precision determination of \mbmz\ from \znull\ boson decays combined
with improved determinations of \mbmh\ from Higgs boson decays would
constitute a new and stringent test of the SM. The b quark couples to
the \znull\ via the electroweak interaction and the observed mass is
related to the strong interaction while the Higgs boson couples to the
b quark directly via its Yukawa coupling. It will be interesting to
see such a test become reality with data from future experimental
facilities.



\end{document}